\RequirePackage{fix-cm}
\documentclass[smallextended]{svjour3}       
\smartqed  
\usepackage{graphicx}
\usepackage{amsmath, amssymb, verbatim}

\newtheorem{remark1}[theorem]{Remark}

\newtheorem{definition}[theorem]{Definition}
\newtheorem{proposition}[theorem]{Proposition}

\newtheorem{example1}[theorem]{Example}

\newenvironment{remark}{\begin{remark1}\rm}{\end{remark1}}
\newenvironment{example}{\begin{example1}\rm}{\end{example1}}

\numberwithin{equation}{section}
\numberwithin{theorem}{section}

\newcommand{\of}[1]{\ensuremath{\left( #1 \right)}}

\newcommand{\abs}[1]{\ensuremath{\left| #1 \right|}}
\newcommand{\cb}[1]{\ensuremath{ \left\{ #1 \right\} }}
\newcommand{\sqb}[1]{\ensuremath{ \left[ #1 \right] }}

\newcommand{\bs}{\backslash}

\newcommand{\vp}{\ensuremath{\varphi}}

\newcommand{\pend}{\hfill $\square$}

\newcommand{\R}{\mathrm{I\negthinspace R}}

\newcommand{\E}{\mathrm{I\negthinspace E}}

\newcommand{\One}{\mathrm{1\negthickspace I}}

\newcommand{\co}{{\rm co \,}}

\newcommand{\diag}{{\rm diag}}

\journalname{Mathematics and Financial Economics}
\begin{document}

\title{Set-valued average value at risk and its computation\thanks{Birgit Rudloff's research was supported by NSF award DMS-1007938.}
}


\author{Andreas H. Hamel         \and
        Birgit Rudloff  \and
        Mihaela Yankova
}

\authorrunning{A. Hamel, B. Rudloff, M. Yankova} 

\institute{Andreas H. Hamel \at
              Yeshiva University\\ Department of Mathematical Sciences\\
              \email{hamel@yu.edu}           
           \and
          Birgit Rudloff \at
              Princeton University\\ ORFE, BCF\\tel. + 1(609)258-4558\\
fax + 1(609)258-3791\\
               \email{brudloff@princeton.edu}
          \and
          Mihaela Yankova \at
              Barclays Capital, New York\\ \email{mihaela.yankova@gmail.com}
}

\date{Received: date / Accepted: date}

\maketitle

\begin{abstract}
New versions of the set-valued average value at risk  for multivariate risks are introduced by generalizing the well-known certainty equivalent representation to the set-valued case. The first 'regulator' version is independent from any market model whereas the second version, called the market extension, takes trading opportunities into account. Essential properties of both versions are proven and an algorithmic approach is provided which admits to compute the values of both version over finite probability spaces. Several examples illustrate various features of the theoretical constructions.
\end{abstract}

\section{Introduction}

The average value at risk for univariate random variables may be seen as a prototype for sublinear (coherent) risk measures. It has several representations and many remarkable properties like law invariance and the Fatou property, and it is linked with second order stochastic dominance. For this and more details, compare \cite[Chap. 4]{FoellmerSchied11}, \cite{AcerbiTasche02,Pflug00}.

Moreover, the computation of its values is a tractable problem: Rockafellar and Uryasev \cite{RockafellarUryasev00} discovered that the computation and optimization of the scalar AV@R over a finite probability space can be done via linear programming methods.

In this note, we introduce new set-valued versions of the average value at risk for multivariate random variables, show essential properties and provide a computational method to obtain their values. The first version is called 'regulator average value at risk' since it does not take into account trading opportunities while the second version is called 'the market extension' since it involves a specific market model.

We will consider market models with proportional transaction costs/bid-ask spreads. It has been shown in \cite{JouiniMeddebTouzi04}  that for such models, set-valued risk measures are an appropriate instrument to evaluate multivariate risks. Later,  a more complete theory for such risk measures has been formulated in \cite{HamelHeyde10}, \cite{HamelHeydeRudloff11}.

We also show that the regulator version of the set-valued average value at risk can be understood as a set-valued representation of a non-complete risk preference. In turn, this preference is represented as the intersection of a collection of complete risk preferences which correspond to scalarizations of the set-valued risk measure. Such a representation is useful for computational issues since it involves nothing else than a family of well-known scalar average value at risks. It also is helpful from a conceptual point of view since it provides a link between set-valued risk measures and existing concepts for scalar risk measures for multivariate positions as defined, for example, in \cite{EkelandSchachermayer11}.

One may ask the question why (seemingly complicated) set-valued functions are used as risk measures. The idea is to collect all deterministic initial portfolios which compensate for the risk of a multivariate position into one set and call this set the value of the risk measure at this position. Of course, this set, in general, includes many elements which "super-compensate" the risk, and a decision maker would only be interested in some minimal elements of this set. The point is that there are, again in general, many non-comparable minimal elements if transaction costs are present. 

For mathematical reasons, it is much easier to work with the whole set of super-compensating portfolios rather than with the set of its minimal elements (a set, nontheless!): Important properties like convexity are easily obtained for the whole set and almost never for the set of minimal elements. Surprisingly, the numerical method presented in this note provides the knowledge about the whole set in terms of (some of its) minimal elements with respect to a given pre-order, so the decision maker gets precisely the information (s)he may need.

The examples at the end illustrate various features of the theoretical construction, among them the geometry of the image sets. Moreover, it is shown that one can also obtain information about trading strategies (example \ref{ex_AVAR_mar_d=2}) and that a scalar approach via liquidation has clear disadvantages compared to the set-valued one (see remark \ref{remark ex_set_vs_scalar}).  Moreover, we show how one can deal, at least in principle, with interest rates (example  \ref{ex_d=5,m=2,r>0}) and exchange rate risks (example \ref{ExAVaRReg_d=3}).

This note utilizes the general framework given in \cite{HamelHeydeRudloff11}, and the results are partially based on \cite{YankovaSeniorThesis11}.

\section{Set-valued average value at risk}

\subsection{The regulator case}

Let $d \geq 1$ be a positive integer and $\of{\Omega, \mathcal{F}_T, P}$ a probability space. A multivariate random variable is an $\mathcal{F}_T$-measurable function $X \colon \Omega \to \R^d$ for $d \geq 2$. If $d=1$, the random variable is called univariate. Denote by $L^0_d = L^0_d\of{\Omega, \mathcal{F}_T, P}$ the linear space of the equivalence classes (with respect to the probability measure $P$) of $\R^d$-valued random variables.  An element $X \in L^0_d$ has components $X_1, \ldots, X_d$ in $L^0 = L^0_1$. The symbol $\One$ denotes the random variable in $L^0$ which has $P$-almost surely the value $1$. Finally, $L^1_d = L^1_d\of{\Omega, \mathcal{F}_T, P}$ denotes the linear space of all $X\in L^0_d$ with $\int_\Omega \abs{X\of{\omega}} \, dP < +\infty$, where $\abs{\cdot}$ stands for an arbitrary, but fixed norm on $\R^d$. 

We shall write
\[
\of{L^0_d}_+ = \cb{X \in L^0_d \mid P\of{\cb{\omega \in \Omega \mid X\of{\omega} \in \R^d_+}} = 1}
\]
for the set of $\R^d$-valued random variables with $P$-almost surely non-negative components, and define $\of{L^1_d}_+=L^1_d \cap \of{L^0_d}_+$. The set $\of{L^1_d}_+$ is a closed convex cone in the Banach space $L^1_d$. If $d=1$ we write $L^0_+$ and $L^1_+$ for $\of{L^0_d}_+$ and $\of{L^1_d}_+$, respectively.

The following definition gives a primal representation for the set-valued average value at risk which extends the certainty equivalent representation of the scalar $AV@R$ as given in \cite{RockafellarUryasev00}. It involves a linear subspace $M \subseteq \R^d$, called the space of eligible assets. The idea is that a risk evaluation of a multivariate position could be done in terms of portfolio vectors instead of units of a single num\'{e}raire. This is motivated by the fact that in the presence of transaction cost there exist portfolios which cannot be exchanged into one another. A natural choice is $M = \R^m \times \cb{0}^{d-m}$, $1 \leq m \leq d$, i.e, the first $m$ out of $d$ assets are eligible as deposits (as in \cite{JouiniMeddebTouzi04}, \cite{ArtDelKochMed09}), but more general constructions are possible (see \cite{HamelHeydeRudloff11} and example \ref{ExAVaRReg_d=3} below). Let us denote $M_+ = M \cap \R^d_+$ and assume $M_+$ is non-trivial, i.e. $M_+ \neq \cb{0}$.

The reader may keep in mind that the multivariate position $X$ models a random future portfolio consisting of $d$ assets where each component gives the number of units of the corresponding asset hold at the future time. Thus, $X_i\of{\omega}$ is the number of units of asset $i$ in the portfolio if $\omega$ occurs. Thus, we follow the "physical unit" approach initiated by Kabanov \cite{K99}. The same interpretation applies to a risk compensating portfolio $u \in M$ deposited at initial time: $u_i$ is the number of units of asset $i$ potentially given as a deposit at initial time.

We will use the symbol $\diag\of{\alpha}$ with $\alpha \in \R^d$ in order to denote the $d \times d$ matrix with the components of the vector $\alpha$ as entries of its main diagonal and zero entries elsewhere.

\begin{definition} \label{DefRegAVaR} Let $\alpha \in (0,1]^d$ and $X \in L^0_d$. The regulator average value at risk at $X$ is defined as
\begin{equation}\label{def_AVAR_reg}
AV@R^{reg}_\alpha\of{X} = 
  \cb{\diag\of{\alpha}^{-1}\E\sqb{Z} - z \mid
    Z \in \of{L^1_d}_+, \; X + Z - z\One \in \of{L^0_d}_+, \; z \in \R^d} \cap M.
\end{equation}
\end{definition}

\begin{remark} 
\label{RemScalar} 
If $m = d = 1$, the two conditions $Z \in L^1_+$ and $X + Z - z\One \in L^0_+$ are equivalent to $Z \geq \of{-X + z\One}^+$ with $X^+ = \max\cb{0, X}$ ($P$--a.s.). We obtain $AV@R^{reg}_\alpha\of{X} = AV@R^{sca}_\alpha\of{X} + \R_+$ with
\begin{equation}
\label{Eq AV@R1}
  AV@R^{sca}_\alpha\of{X} = \inf_{z \in \R} \cb{\frac{1}{\alpha}\E\sqb{\of{-X + z\One}^+} - z},
\end{equation}
which is the optimized certainty equivalent representation of the $AV@R$ found by Rockafellar and Uryasev \cite{RockafellarUryasev00}. Compare also F\"ollmer and Schied \cite{FoellmerSchied11}, formula~(4.42). Note that $AV@R^{sca}_\alpha\of{X} = +\infty$ if $\of{-X}^+ \not\in L^1$.
\end{remark}

\begin{remark}
\label{RemInterpretationAVaRReg}
The following observations will lead to an interpretation of formula \eqref{def_AVAR_reg}. Let $M = \R^m \times \cb{0}^{d-m}$ (hence $M_+ = \R^m_+ \times  \cb{0}^{d-m}$) with $1 \leq m \leq d$, i.e., the first $m$ assets are eligible. In this case, the objective and the constraints can be considered component-wise, and we obtain for $i =1, \ldots, m$
\begin{align*}
& \cb{\frac{1}{\alpha_i}\E\sqb{Z_i} - z_i \mid Z_i \in L^1_+, \; X_i + Z_i - z_i\One \in L^0_+, \; z_i \in \R} \\
	& \quad = \cb{\frac{1}{\alpha_i}\E\sqb{Z_i} - z_i \mid Z_i \geq \max\cb{0, z_i\One - X_i} = \of{z_i\One - X_i}^+,  \; z_i \in \R} \\
	& \quad= \inf\cb{\frac{1}{\alpha_i}\E\sqb{\of{z_i\One - X_i}^+} -z_i \mid z_i \in \R} + \R_+ = AV@R^{sca}_{\alpha_i}\of{X_i} + \R_+.
\end{align*}
For $i = m+1, \ldots, d$, there must exist $Z_i \in L^1_+$, $z_i \in \R^d$ such that $0 = \frac{1}{\alpha_i}\E\sqb{Z_i} - z_i$ and $X_i + Z_i - z_i\One \in L^0_+$. This means $AV@R^{sca}_{\alpha_i}\of{X_i}  \leq 0$. Altogether, the set-valued regulator AV@R produces the component-wise scalar AV@R for the first $m$ components plus the cone $\R^m_+ \times \cb{0}^{d-m}$ under the constraint that the scalar AV@R for the last $d-m$ components is at most zero. If the latter is not the case, $AV@R^{reg}_\alpha\of{X} = \emptyset$. So, the set-valued regulator average value at risk is either "point plus cone" or the empty set. This simple geometry will change if $M$ has a different structure as shown in example \ref{ExAVaRReg_d=3}.
\end{remark}

\begin{proposition}
\label{PropAVaR}
The function $X \mapsto AV@R^{reg}_\alpha\of{X}$ enjoys the following properties:

(a) it is positively homogeneous, i.e.
\[
\forall X \in L^0_d, \forall s> 0 \colon AV@R^{reg}_\alpha\of{sX} = sAV@R^{reg}_\alpha\of{X},
\]

(b) it is subadditive, i.e.
\[
\forall X^1, X^2 \in L^0_d \colon 
	AV@R^{reg}_\alpha\of{X^1 + X^2} \supseteq AV@R^{reg}_\alpha\of{X^1} + AV@R^{reg}_\alpha\of{X^2},
\]

(c) it is $M$-translative, i.e.
\[
\forall X \in L^0_d, \forall u \in M \colon AV@R^{reg}_\alpha\of{X + u\One} = AV@R^{reg}_\alpha\of{X} - u,
\]

(d) it is monotone with respect to $\of{L^0_d}_+$, i.e.
\[
X^1, X^2 \in L^0_d, \; X^2 - X^1 \in \of{L^0_d}_+ \quad \Rightarrow \quad 
	AV@R^{reg}_\alpha\of{X^2} \supseteq AV@R^{reg}_\alpha\of{X^1},
\]

(e) for each $X \in L^0_d$, the set $AV@R^{reg}_\alpha\of{X} \subseteq M$ is convex and satisfies
\[
AV@R^{reg}_\alpha\of{X} + M_+ = AV@R^{reg}_\alpha\of{X}.
\]

In particular, $AV@R^{reg}_\alpha\of{0}$ is a convex cone which satisfies
\[
M_+ \subseteq AV@R^{reg}_\alpha\of{0}, \quad AV@R^{reg}_\alpha\of{0} \cap -M_+ = \cb{0}.
\]
\end{proposition}

{\sc Proof.} (a) Easy to check. (b) Taking $X_1, X_2 \in L^0_d$ we obtain
\begin{align*} 
& AV@R_{\alpha}^{reg}\of{X_1}+ AV@R_{\alpha}^{reg}\of{X_2} \\
& = \left\{\diag\of{\alpha}^{-1} \E\sqb{Z^1} - z^{1} + \diag\of{\alpha}^{-1} \E\sqb{Z^{2}} - z^{2} \mid
   Z^1, Z^2 \in \of{L^1_d}_+, \; z^1, z^2 \in \R^d, \right.\nonumber\\
& \qquad\qquad\qquad \left. X_1 + Z^1 - z^1\One \in  \of{L^0_d}_+,  \; X_2 + Z^2 - z^2\One \in \of{L^0_d}_+\right\} \cap M \\
& \subseteq \left\{\diag\of{\alpha}^{-1} \E\sqb{Z^1+Z^2} - \of{z^1 + z^2} \mid
    Z^1 + Z^2 \in \of{L^1_d}_+, \right.\nonumber\\
& \qquad\qquad\qquad \left. X_1 + X_2 + Z^1 + Z^2 - \of{z^1 + z^2}\One \in \of{L^0_d}_+, \; z^1+z^2 \in \R^d \right\} \cap M \\
& = \cb{\diag\of{\alpha}^{-1} \E\sqb{Z} - z \mid
    Z \in \of{L^1_d}_+, \; X_1 + X_2 + Z - z\One \in \of{L^0_d}_+, \; z \in \R^d } \cap M \\
& = AV@R_{\alpha}^{reg}\of{X_1+X_2}.
\end{align*}
Again, (c) is straightforward. (d) $AV@R^{reg}_\alpha$ is $\of{L^0_d}_+$-monotone since for $U \in \of{L^0_d}_+$ we have $U + \of{L^0_d}_+\subseteq \of{L^0_d}_+$ and therefore
\begin{align*}
 AV@R_{\alpha}^{reg}\of{X-U} 
  &  = \cb{\diag\of{\alpha}^{-1} \E\sqb{Z} - z \mid
    Z \in \of{L^1_d}_+, \; X+ Z - z\One \in U + \of{L^0_d}_+, \; z \in \R^d} \cap M\\
   & \subseteq  AV@R_{\alpha}^{reg}\of{X}.
\end{align*}
(e) The convexity of $AV@R_{\alpha}^{reg}\of{X}$ is obtained from (a) and (b) since for $s \in \of{0, 1}$
\[
AV@R_{\alpha}^{reg}\of{X} = AV@R_{\alpha}^{reg}\of{sX + \of{1-s}X} \supseteq 
	sAV@R_{\alpha}^{reg}\of{X} + \of{1-s}AV@R_{\alpha}^{reg}\of{X}.
\]
(a) yields that $AV@R_{\alpha}^{reg}\of{0}$ is a convex cone. If $k \in M_+$ then $\of{L^0_d}_+ +k\One \subseteq\of{L^0_d}_+$ and hence
\begin{align*}
&AV@R_{\alpha}^{reg}\of{X} + k  
 =\cb{\diag\of{\alpha}^{-1}\E\sqb{Z} - \tilde{z} + k \mid
    Z \in \of{L^1_d}_+, \; X + Z - \tilde{z}\One \in \of{L^0_d}_+, \; \tilde{z} \in \R^d} \cap M \\
& = \cb{\diag\of{\alpha}^{-1}\E\sqb{Z} - z \mid
    Z \in \of{L^1_d}_+, \; X + Z - z\One \in k \One +\of{L^0_d}_+, \; z \in \R^d} \cap M 
 \subseteq  AV@R_{\alpha}^{reg}\of{X}.
\end{align*}
Finally, while obviously $0 \in AV@R^{reg}_\alpha\of{0} \neq \emptyset$, take $k \in M_+\bs\cb{0}$ (non-empty by assumption) and $s>0$. Assume $-sk \in AV@R_{\alpha}^{reg}\of{0}$. Then, there is $Z \in \of{L^1_d}_+$, $z \in \R^d$ such that $Z - z\One \in \of{L^0_d}_+$ and $-sk = \diag\of{\alpha}^{-1}\E\sqb{Z} - z$. These conditions imply $Z \in \of{L^1_d}_+$ and
\[
Z - sk - \diag\of{\alpha}^{-1}\E\sqb{Z} \in \of{L^1_d}_+.
\]
At least one component of $k$ is positive, say $k_i > 0$. Multiplying the above inclusion with $Y \in \of{L^\infty_d}_+$ defined by $Y_i = e^i\One$ and $Y_j \equiv 0$ for $j \neq i$ and taking the expected value we obtain
\[
\E\sqb{Z_i}\of{1 - \frac{1}{\alpha_i}} -sk_i \geq 0.
\]
Since $\alpha_i \leq 1$ and  $\E\sqb{Z_i} \geq 0$, this contradicts  $sk_i > 0$. Therefore, $-sk \not\in AV@R_{\alpha}^{reg}\of{0}$ for $s>0$ and $AV@R_{\alpha}^{reg}\of{0}\neq M$. \pend

\begin{remark}
\label{RemAcceptanceSetReg}
The acceptance set of a set-valued risk measure is the set of all $X \in L^0_d$ such that $0 \in M$ is included in the value of the risk measure at $X$ (see \cite{HamelHeydeRudloff11}), and there is the usual one-to-one relationship between acceptance sets and risk measures.  We obtain from definition \ref{DefRegAVaR} that the acceptance set of $AV@R^{reg}_\alpha$ is
\begin{align*}
\mathcal A^{reg} & = \cb{X \in L^0_d \mid 0 \in AV@R^{reg}_\alpha\of{X}} \\
	& = \cb{X \in L^0_d \mid \exists Z \in \of{L^1_d}_+ \colon
	X + Z - \diag\of{\alpha}^{-1}\E\sqb{Z} \One \in \of{L^0_d}_+}. 
\end{align*}
The set $\mathcal A^{reg}$ obviously is a convex cone, and it satisfies $\mathcal A^{reg} +  \of{L^0_d}_+ = \mathcal A^{reg}$ as well as the following two non-triviality conditions: $M_+\One \subseteq \mathcal A^{reg}$ and $\of{-M_+\One} \cap \mathcal A^{reg} = \cb{0}$. Indeed, while the first condition is trivial (choose $Z = 0$), the second follows from (c) and (e) of proposition \ref{PropAVaR}. The average value at risk can be recovered from $\mathcal A^{reg}$ by
\[
AV@R^{reg}_\alpha\of{X} = \cb{u \in M \mid X + u\One \in \mathcal A^{reg}}, 
\]
that is, it includes all initial eligible portfolios which make the position $X$ acceptable.
\end{remark}

\subsection{Scalarization and interpretation}

In the following, we assume that the value $AV@R^{reg}_\alpha(X)$ is a non-empty closed set. Since it is also convex, it is the intersection of all closed half-space including it. Such a half space has a normal $w \in (\R^d_+ \cap M)^+ \bs\cb{0} = (\R^d_+ + M^\perp)\bs\cb{0}$. Thus,
\[
u \in AV@R^{reg}_\alpha\of{X} \; \Leftrightarrow \; \forall w \in (\R^d_+ + M^\perp)\bs\cb{0} \colon w^Tu \geq \inf\cb{w^Tx \mid x \in AV@R^{reg}_\alpha\of{X}}.
\]
Now, consider the extended real-valued function
\[
X \mapsto \vp_{AV@R^{reg}_\alpha, w}\of{X} = \inf\cb{w^Tu \mid u \in AV@R^{reg}_\alpha\of{X}}.
\]
Note that $\vp_{AV@R^{reg}_\alpha, w + v} = \vp_{AV@R^{reg}_\alpha, w}$ whenever $v \in M^\perp$ since $AV@R^{reg}_\alpha\of{X} \subseteq M$ for all $X \in L^0_d$. Thus, one can restrict the functions $\vp_{AV@R^{reg}_\alpha, w}$ to those with $w \in \R^d_+ $ and obtains
\[
u \in AV@R^{reg}_\alpha\of{X} \; \Leftrightarrow \; \forall w \in \R^d_+, \; \forall v \in M^\perp, \; w+v \neq 0 \colon \of{w + v}^Tu \geq \vp_{AV@R^{reg}_\alpha, w}\of{X}.
\]
Defining the set $B\of{1} = \{w \in \R^d_+ \mid \sum_{i=1}^d w_i = 1\}$ and using the positive homogeneity with respect to $w$ of the functions $\vp_{AV@R^{reg}_\alpha, w}$ one can further show that
\begin{multline*}
u \in AV@R^{reg}_\alpha\of{X} \; \Leftrightarrow \;
\forall w \in  B\of{1}, \; \forall v \in M^\perp, \; w+v \neq 0 \colon \of{w + v}^Tu \geq \vp_{AV@R^{reg}_\alpha, w}\of{X}.
\end{multline*}

It turns out that the functions $\vp_{AV@R^{reg}_\alpha, w}$ are composed of scalar AV@R-type functions for the components of $X$. More precisely, we have the following result.

\begin{proposition}
\label{PropScalarization}
Let $AV@R^{reg}_\alpha\of{X} \neq \emptyset$, $w \in \R^d_+\bs\cb{0}$ and $I_>\of{w} = \cb{i \in \cb{1, \ldots, d} \mid w_i > 0}$. Then
\begin{equation}
\label{EqAVaRMix}
\vp_{AV@R^{reg}_\alpha, w}\of{X} = \sum\limits_{i \in I_>\of{w}} w_iAV@R^{sca}_{\alpha_i}\of{X_i}.	
\end{equation}
\end{proposition}

{\sc Proof.} The crucial observation is as follows. If $u \in AV@R^{reg}_\alpha\of{X}$ then there are $Z \in \of{L^1_d}_+$, $z \in \R^d$ such that $u = \diag\of{\alpha}^{-1}\E\sqb{Z} - z$ and $X + Z - z\One \in \of{L^0_d}_+$. Hence
\[
w^Tu  = \sum_{i \in I_>\of{w}} \of{\frac{1}{\alpha_i}\E\sqb{w_iZ_i} -w_iz_i}
\]
since there is zero contribution to this sum whenever $w_i = 0$. Since the components are "separated" we have
\begin{align*}
\vp_{AV@R^{reg}_\alpha, w}\of{X} 
	& = \sum\limits_{i \in I_>\of{w}} \inf\cb{\of{\frac{1}{\alpha_i}\E\sqb{w_iZ_i} -w_iz_i} \mid Z_i \in \of{L^1}_+, X_i + Z_i - z_i\One \geq 0} \\
	& = \sum\limits_{i \in I_>\of{w}} \inf\cb{\of{\frac{1}{\alpha_i}\E\sqb{w_iZ_i} -w_iz_i} \mid w_iZ_i \in \of{L^1}_+, w_iX_i + w_iZ_i - w_iz_i\One \geq 0} \\
	& = \sum\limits_{i \in I_>\of{w}} \inf\cb{\of{\frac{1}{\alpha_i}\E\sqb{U_i} -r_i} \mid U_i \in \of{L^1}_+, w_iX_i + U_i - r_i\One \geq 0} 
\end{align*}
which, in view of remark \ref{RemScalar} and positive homogeneity of the scalar average value at risk, already is the result. \pend

\medskip Proposition \ref{PropScalarization} together with the discussion preceding it provides the following representation of the set-valued average value at risk. If $AV@R^{reg}_\alpha\of{X}$ is closed then

\begin{equation}
\label{EqAVaRSetified}
AV@R^{reg}_\alpha\of{X} = \bigcap_{w \in  B\of{1}, \, v \in M^\perp, \, w+v \neq 0}\cb{u \in \R^d \mid \of{w + v}^Tu \geq \vp_{AV@R^{reg}_\alpha, w}\of{X}}.
\end{equation}

Thus, formula \eqref{EqAVaRMix} means that the functions $\vp_{AV@R^{reg}_\alpha, w}$ entering \eqref{EqAVaRSetified} are mixtures (= convex combinations) of the scalar average value at risks applied to the components of the multi-variate position $X$. Moreover, $u \in AV@R^{reg}_\alpha\of{X}$ if and only if the mixture $w^Tu = \of{w + v}^Tu$ (since $u \in M$) compensates for the risk of the mixture $w^TX$ for all $w \in B\of{1}$. Put differently, the "mixed" number $w^Tu$ is not less than the corresponding mixture of the scalar average value at risks applied to the components of $X$. This interpretation also explains why the vector $\alpha$ may have different components. The decision maker may have a different attitude towards the risk of each component of the portfolio vector $X$ - which contains the "physical" units of each asset.

Finally, the vector $w \in  B\of{1}$ may be understood as a weighting of the $d$ assets: A decision maker may have more and less favorable assets when it comes to risk evaluation. The formulas \eqref{EqAVaRMix} and \eqref{EqAVaRSetified} also mean that the regulator does not care about weightings of the decision makers: The regulator needs to be on the safe side no matter what kind of weighting a particular decision maker may have.

The reader may observe that this discussion shows that there is a relationship between the scalar average value at risk applied to the components of $X$ and the set-valued one. This relationship does not mean, in general, that the set-valued AV@R can always be represented as "vector of the component-wise taken scalar AV@Rs plus a convex cone". Example \ref{ExAVaRReg_d=3} provides a counterexample.

Rather, the relationship expressed in \eqref{EqAVaRSetified} may be interpreted as follows. Each scalar function $\vp_{AV@R^{reg}_\alpha, w}$ induces a complete risk preference whereas the risk preference represented by the set-valued function $AV@R^{reg}_\alpha$ is a non-complete one: There might be positions $X^1, X^2$ whose risks cannot be compared.

We conclude this discussion with the following result which relates the scalarizations $\vp_{AV@R^{reg}_\alpha, w}$ to existing concepts.

\begin{proposition}
\label{PropScalarTranslative}
Let $w \in  B\of{1}$ and $e = \of{1, \ldots, 1}^T \in \R^d$. Then
\[
\forall s \in \R, \; \forall X \in L^0_d \colon\; \vp_{AV@R^{reg}_\alpha, w}\of{X + se\One} = \vp_{AV@R^{reg}_\alpha, w}\of{X} - s.
\]
\end{proposition}

{\sc Proof.} Immediate using \eqref{EqAVaRMix} and the translation property of the scalar average value at risk. \pend

\medskip We note that proposition \ref{PropScalarTranslative} shows that the functions $\vp_{AV@R^{reg}_\alpha, w}$ share all properties of scalar risk measures for $\R^d$-valued risks as defined in \cite[Definition 1.2]{EkelandSchachermayer11}, and this property does not coincide in general with the translation property {\em M2)} in \cite[Definition 3.1]{BurgertRueschendorf06}.

\subsection{The market extension}

\medskip The average value at risk from definition \ref{DefRegAVaR} does not take into account a market model. Therefore, we define its market extension (see \cite{HamelHeydeRudloff11} for a general definition and further motivation). We consider a discrete market at time points $\cb{0,1,...,T}$ and a filtered probability space $(\Omega,\mathcal{F}_T,\of{\mathcal{F}_t}_{t=0}^T,P)$ satisfying the usual conditions.  A discrete conical market model is a sequence of $\mathcal F_t$-measurable functions $K_t \colon \Omega \to 2^{\R^d}$, $t=0,1, \ldots,T$ with $\R^d_+ \subseteq K_t \neq \R^d$ such that $K_t\of{\omega}$ is a closed convex cone for each $\omega \in \Omega$ and all $t \in \cb{0, \ldots, T}$. These cones are called solvency cones and appear, for example, when proportional transaction costs are present. Compare \cite{K99,S04,KS09}. Measurability of the solvency cone mapping means, as usual, that 
for each open set $B \subseteq \R^d$ the set $\cb{\omega \in \Omega \mid K_t\of{\omega} \cap B \neq \emptyset}$ is an element of $\mathcal{F}_t$.

We denote $K^M_0 = K_0 \cap M$ which is non-trivial since $M_+$ is non-trivial and $\R^d_+ \subseteq K_0$.
Let $L^0_d\of{\Omega, \mathcal{F}_t, P}$ be the linear space of equivalence classes of $\R^d$-valued, $\mathcal{F}_t$-measurable random variables. Further, denote
\[
L^0_d\of{K_t} = \cb{X \in L^0_d\of{\Omega, \mathcal{F}_t, P} \mid P\of{\cb{\omega \in \Omega \mid X\of{\omega} \in K_t\of{\omega}}} = 1}
\]
for $t \in \cb{0, 1, \ldots, T}$. Finally, define the set
\[
C_T = -\sum_{t = 0}^T L^0_d\of{K_t}\subseteq L^0_d
\]
which includes all payoffs which can be generated by trading with initial portfolio $0 \in \R^d$. Elements of $C_T$ are also called freely available payoffs, and $C_T$ is a convex cone in $L^0_d$.

\begin{definition}
\label{DefAVaRMarket}
Let $\alpha \in (0,1]^d$ and $X \in L^0_d$. The market extension of $AV@R^{reg}_\alpha$ at $X$ is defined as 
\begin{equation}
\label{EqAVaRMarExt}
 AV@R^{mar}_\alpha\of{X} 
  = \cb{\diag\of{\alpha}^{-1}\E\sqb{Z} - z \mid
    Z \in \of{L^1_d}_+, \; X + Z - z\One \in -C_T, \; z \in \R^d} \cap M.
\end{equation}
\end{definition}

\begin{remark}
\label{RemMarExt1}
Formula \eqref{EqAVaRMarExt} can be rewritten as 
\begin{equation}
\label{EqMarExtInf}
AV@R^{mar}_\alpha\of{X} = \bigcup\cb{AV@R^{reg}_\alpha\of{X + Y} \mid Y \in C_T}.
\end{equation}
Indeed, this follows from the equation $-C_T + \of{L^0_d}_+  = -C_T$ and the definition of $AV@R^{reg}_\alpha$. Formula \eqref{EqMarExtInf} has a nice financial interpretation: The market AV@R produces all initial portfolios which compensate for the risk of $X$ plus a freely available payoff. Thus, an investor who is faced with the risk described by $X$ may add a freely available payoff to $X$ and then compensate for the overall risk. This offers greater flexibility once a market model is fixed.
\end{remark}

\begin{remark}
\label{RemMarExt2}
Formula \eqref{EqMarExtInf} has another, mathematical interpretation. Define the set
\[
\mathcal C\of{M, M_+} = \cb{D \subseteq M \mid D = \co\of{D + M_+}}.
\]
Then $\of{\mathcal C\of{M, M_+}, \supseteq}$ is an order complete lattice in the sense that every subset $\mathcal D \subseteq \mathcal C\of{M, M_+}$ has an infimum and a supremum in $\mathcal C\of{M, M_+}$ with respect to $\supseteq$, namely
\[
\inf \mathcal D = \co\bigcup_{D \in \mathcal D} D \quad \text{and} \quad 
	\sup \mathcal D = \bigcap_{D \in \mathcal D} D,
\]
respectively. Then,
\begin{equation}
\label{EqAVaRInfRep}
AV@R^{mar}_\alpha\of{X} = \inf\cb{AV@R^{reg}_\alpha\of{X + Y} \mid Y \in C_T}
\end{equation}
where the infimum is taken in $\of{\mathcal C\of{M, M_+}, \supseteq}$. Indeed, everything is obvious but the missing convex hull. It can be dropped since the set  $\bigcup\cb{AV@R^{reg}_\alpha\of{X + Y} \mid Y \in C_T}$ is convex due to the convexity of $X \mapsto AV@R^{reg}_\alpha\of{X}$. Indeed, if $u^1, u^2 \in \bigcup\cb{AV@R^{reg}_\alpha\of{X + Y} \mid Y \in C_T}$, then there are $Y^1, Y^2 \in C_T$ such that $u^i \in AV@R^{reg}_\alpha\of{X + Y^i}$, $i=1,2$, and
\begin{multline}
su^1 + \of{1-s}u^2 \in s AV@R^{reg}_\alpha\of{X + Y^1} + \of{1-s}AV@R^{reg}_\alpha\of{X + Y^2} \\
	 \subseteq AV@R^{reg}_\alpha\of{X +sY^1 + \of{1-s}Y^2}
	\subseteq AV@R^{mar}_\alpha\of{X}
\end{multline}
since $sY^1 + \of{1-s}Y^2 \in C_T$ by convexity of $C_T$.

Thus, $AV@R^{mar}_\alpha\of{X}$ is the optimal value of a minimization problem with a set-valued objective.
\end{remark}

\begin{remark}
Another motivation to consider market extensions is given by its economic interpretation as valuation
bounds: The two functions $X \mapsto  AV@R^{mar}_\alpha\of{-X} $ and $X \mapsto -AV@R^{mar}_\alpha\of{X} $ are the set-valued analogs to good deal bounds with respect to the risk measure $AV@R$, see e.g. \cite{JaschkeKuechler2001} for the scalar case. This allows to consider tighter valuation bounds than the ones obtained by superhedging in markets with transaction costs. One can see this by noting that $AV@R^{reg}_\alpha\of{X}\supseteq WC\of{X}$ for all $X\in L^0_d$,
where $WC\of{X}=\{u \in M \mid X + u\One \in \of{L^0_d}_+  \}$ is the worst case risk measure, see section~5.3 in \cite{HamelHeydeRudloff11}. Then, equation  \eqref{EqMarExtInf}  implies
\begin{align*}
AV@R^{mar}_\alpha\of{-X} & \supseteq\bigcup\cb{WC\of{-X + Y} \mid Y \in C_T} = \cb{u \in M \mid -X + u\One \in C_T}. 
\end{align*}
where the right hand side coincides with the set of portfolios in $M$ that allow to superhedge $X$, see \cite{K99,S04,KS09,LR11}. 
Thus, the upper valuation bound $AV@R^{mar}_\alpha\of{-X} \in \mathcal C\of{M, M_+} $ contains more and thus cheaper portfolios compared to the set of superhedging portfolios  and the lower valuation bound $-AV@R^{mar}_\alpha\of{X} \in \mathcal C\of{M, -M_+} $ contains more and thus more expensive portfolios compared to the set of  subhedging portfolios.
\end{remark}

The market extension shares several, but not all properties with the regulator AV@R.

\begin{proposition}
\label{PropAVaRExt}
The market extension $AV@R^{mar}_\alpha$ satisfies (a), (b), (c) of proposition \ref{PropAVaR} and moreover

(d')  it is monotone with respect to $-C_T$, i.e.
\[
X^1, X^2 \in L^0_d, \; X^2 - X^1 \in -C_T \quad \Rightarrow \quad 
	AV@R^{mar}_\alpha\of{X^2} \supseteq AV@R^{mar}_\alpha\of{X^1},
\]

(e') for each $X \in L^0_d$, the set $AV@R^{mar}_\alpha\of{X} \subseteq M$ is convex and satisfies
\[
AV@R^{mar}_\alpha\of{X} + K_0^M = AV@R^{mar}_\alpha\of{X}.
\] 
 
In particular, $AV@R^{mar}_\alpha\of{0}$ is a convex cone with $K^M_0 \subseteq AV@R^{mar}_\alpha\of{0}$.
\end{proposition}

{\sc Proof.} (a), (b), (c) and (e') follow with similar arguments as in proposition \ref{PropAVaR}. (d') If $X^2 - X^1 \in -C_T$ then there is $Y^1 \in C_T$ such that $X^1 = X^2 + Y^1$. Since $C_T$ is a convex cone we have $Y^1 + C_T \subseteq C_T$ and hence by \eqref{EqMarExtInf}
\begin{align*}
AV@R^{mar}_\alpha\of{X^1} & = \bigcup\cb{AV@R^{reg}_\alpha\of{X^2 + Y^1 + Y} \mid Y^1 + Y \in Y^1 + C_T} \\
	& \subseteq \bigcup\cb{AV@R^{reg}_\alpha\of{X^2 + Y^1 + Y} \mid Y^1 + Y \in C_T} \\
	& = \bigcup\cb{AV@R^{reg}_\alpha\of{X^2 + Y^2} \mid Y^2 \in C_T} = AV@R^{mar}_\alpha\of{X^2}.
\end{align*}
This completes the proof. \pend

\medskip Property (e') has the following interpretation. If $X^2 - X^1 \in -C_T$, then $X^1$ is the sum of $X^2$ and a freely available payoff. Unless there is some kind of arbitrage opportunity, this means that $X^1$ is somehow "riskier" than $X^2$, hence the set of initial eligible portfolios which compensate for the risk of $X^1$ should not be "greater" than the corresponding set for $X^2$.

A particular feature of the market extension is that "finiteness" (see (e) of proposition \ref{PropAVaR}) cannot be guaranteed in general. 

\begin{remark}
\label{RemAcceptanceSetMar} The acceptance set of $AV@R^{mar}_\alpha$ is given by
\[
\mathcal A^{mar}  = \cb{X \in L^0_d \mid \exists Z \in \of{L^1_d}_+ \colon
	X + Z - \diag\of{\alpha}^{-1}\E\sqb{Z} \One \in -C_T} = \mathcal A^{reg} - C_T.
\]
\end{remark}

\section{Set-valued AV@R over finite probability spaces}

In the rest of the paper, we impose the following assumptions and notational conventions.

\medskip\noindent (H1) $\abs{\Omega} = N$. $\mathcal{F}_T = 2^\Omega$, where the probability
measure $P$ is given by $N$ strictly positive numbers $p = \of{p_1, p_2, \ldots, p_N}$ with $\sum_{n=1}^N p_n=1$
and $P\of{\cb{\omega_n}}=p_n$, $n = 1, \ldots, N$.

\medskip\noindent (H2) The vectors $b^1, \ldots, b^m \in \R^d$ form a basis of the space $M$ of eligible portfolios, the vectors $b^{m+1}, \ldots, b^d \in \R^d$ form a basis of $M^\perp$. Of course, $1 \leq m \leq d$, and $b^1, \ldots, b^d$ form a basis of $\R^d$. We exclude neither the case $m = 1$ nor the case $m = d$.

\medskip\noindent (H3) The cone $K_0$ is spanned by $h^1, \ldots, h^I \in \R^d$, thus
it is a finitely generated (hence closed) convex cone.

\medskip\noindent (H4) The cone $K^M_0 = K_0 \cap M$ is spanned by $g^1, \ldots, g^L \in \R^d$, thus it also is a finitely generated, closed convex cone. Note that this collection can be entirely different from $h^1, \ldots, h^I$.

\medskip\noindent (H5) For each $\omega \in \Omega$, the cone $K_T\of{\omega}$ is spanned by $k^1\of{\omega}, \ldots,
k^{J\of{\omega}}\of{\omega}$, thus it is a finitely generated (hence closed) convex cone.

\medskip Note that assumptions (H3) and (H5) are always satisfied in markets with proportional transaction costs where the solvency cones are generated by the bid and ask exchange rates between any two of the $d$ assets as for example considered in \cite{K99,S04,KS09}.

\subsection{The discrete version of $AVaR_\alpha^{reg}$}
\label{subsec AVAR_reg}

In the following, we shall reformulate $AV@R^{reg}_\alpha$ given in \eqref{def_AVAR_reg} of definition \ref{def_AVAR_reg} in a linear programming language. We use the representation of the random variables $X, Z \colon \Omega \to \R^d$ by
$x_{in} = X_i\of{\omega_n}$ and $z_{in} = Z_i\of{\omega_n}$ $i = 1, \ldots, d$, $n = 1, \dots, N$ and set
\begin{align*}
\hat{x} = \of{x_{11}, \ldots, x_{d1}, x_{12}, \ldots, x_{dN}}^T \in \R^{dN}\\
\hat{z} = \of{z_{11}, \ldots, z_{d1}, z_{12}, \ldots, z_{dN}}^T \in \R^{dN}.
\end{align*}

First, the condition $Z \in \of{L^1_d}_+$ is equivalent to $\hat z \in \R^{dN}_+$. Next, using
\[
\hat{I}_d =\left(
          \begin{array}{c}
           I_d \\
           \vdots  \\
           I_d
           \end{array}
           \right) \in \R^{dN \times d},
\]
where $I_d$ is the $d \times d$ identity matrix and $\hat t \in \R^{dN}_+$, we can write the condition $Z + X - z\One \in \of{L^0_d}_+$ as
\[
\hat z+ \hat x  - \hat I_d z = \hat t.
\]

The objective function $\of{Z, z} \mapsto \diag(\alpha)^{-1} \E\sqb{Z} - z$ can be
given a matrix form as follows. If we define
\[
P_{\of{n}} = p_nI_d \in \R^{d \times d}, \; n=1, \ldots, N,
\]
then $\hat{P} = \of{P_{\of{1}} \; P_{\of{2}} \; \ldots \; P_{\of{N}}}$ is a $d \times
dN$-matrix and
\[
\diag(\alpha)^{-1} \E\sqb{Z} - z = \diag(\alpha)^{-1}\hat{P}\hat{z} - z.
\]

Finally, the constraint $\diag(\alpha)^{-1} \E\sqb{Z} - z\in M$ can be written as follows. Denote by
\[
B_{\of{d-m}} = \left(
        \begin{array}{ccc}
          b_1^{m+1} & \ldots & b_d^{m+1} \\
           \vdots & \ddots & \vdots \\
           b_1^{d} & \ldots & b_d^{d} \\
        \end{array}
      \right) \in \R^{(d-m) \times d}
\]
the matrix containing the generating vectors of $M^\perp$ as rows. Then, the condition $\diag(\alpha)^{-1} \E\sqb{Z} - z\in M$ is equivalent to $B_{\of{d-m}}(\diag(\alpha)^{-1}\hat{P}\hat{z} - z)=0$.

Altogether, we end up with
\begin{multline}
\label{AVaRRegDiscrete}
AV@R_\alpha^{reg}\of{X} = \big\{\diag(\alpha)^{-1}\hat{P}\hat{z} - z \mid  B_{\of{d-m}}(\diag(\alpha)^{-1}\hat{P}\hat{z} - z)=0,
\\
 \hat z+ \hat x  - \hat I_d z = \hat t,\; \hat{z}\in \R^{dN}_+, \;\hat{t}\in \R^{dN}_+, \; z \in \R^d\big\}.
\end{multline}

This shows that $AV@R_\alpha^{reg}\of{X}$ is the image of a polyhedral set under a linear function mapping $\R^{2dN+d}$ into $M$. In particular, by (H1) - (H5) $AV@R_\alpha^{reg}$ has a closed graph and closed values since all these sets are polyhedral.

How can we compute the sets $AV@R_\alpha^{reg}\of{X}$, i.e. the values of the average value at risk? To do this, one must first answer the question what it means "to know" these sets. Here, we advance the view that we know $AV@R_\alpha^{reg}\of{X}$ if we know its minimal points with respect to the order in $M$ generated by $M_+ = \R^d_+ \cap M$, i.e. points $\bar u \in AV@R_\alpha^{reg}\of{X}$ satisfying
\[
\of{\bar u - M_+} \cap AV@R_\alpha^{reg}\of{X} = \cb{\bar u},
\]
and its directions of recession, that is $k \in M$ satisfying
\[
\forall s>0 \colon AV@R_\alpha^{reg}\of{X} + sk \subseteq AV@R_\alpha^{reg}\of{X},
\]
which together form the recession cone. Moreover, in most cases it is sufficient to know those minimal points of $AV@R_\alpha^{reg}\of{X}$ which are also extremal, i.e. vertices. The others can be recovered by taking convex combinations. The set of extremal minimal points is finite since $AV@R_\alpha^{reg}\of{X}$ is polyhedral. This point of view admits a financial as well as a mathematical interpretation. Financially, the minimal points of $AV@R_\alpha^{reg}\of{X}$ represent the eligible portfolios which an investor might want to deposit since all others are more expensive. Mathematically, the set $AV@R_\alpha^{reg}\of{X}$ can be reconstructed by taking the convex hull of the union of the sets $\bar u + M_+$ and adding the recession cone where $\bar u$ runs through the extremal minimal points of $AV@R_\alpha^{reg}\of{X}$, thus taking the infimum in the space $\mathcal C\of{M, M_+}$ (compare remark \ref{RemMarExt2}). This corresponds to a solution concept for set-valued optimization problems like the one in \eqref{EqAVaRInfRep} described in \cite[Chap. 2]{Loehne11Book}.

Thus, one needs an algorithm that computes the (extremal) minimal points and the recession cone of a polyhedral convex set which is  the image of another polyhedral set under a linear map as in \eqref{AVaRRegDiscrete}. Such an algorithm is the one given by H. Benson in \cite{Benson98JGO} which has been extended and modified in \cite{Loehne11Book} and further in the forthcoming \cite{HLR12}.

We emphasize two features of the algorithm\footnote{For actual computations, we used a version of the algorithm, BENSOLVE, which is available online at http://ito.mathematik.uni-halle.de/$\sim$~loehne}: First, computing the minimal points of the set given in \eqref{AVaRRegDiscrete} means to solve a linear vector optimization problem, see, for example, \cite[Part II]{Loehne11Book}. Secondly, Benson's algorithm is perfectly suited for the problems at hand: It basically works in the image space, here $M$ with dimension $m$, which is in most cases very small compared to the dimension of the preimage space, here $2dN+d$.

A few examples shall illustrate the theory and point to some interesting features which appear in the transaction cost setting.

\begin{example}
\label{ex_AVAR_reg_d=2}
The number of assets is $d=2$ and $M = \R^2$, so all initial portfolios are eligible. In a binary model with $N=2$ and $p = \of{0.4, 0.6}$, the payoff is given by
\begin{align*}
X(\omega_1) = \left(
                          \begin{array}{c}
                          12\\
                          -20
                          \end{array}
                        \right), \quad X(\omega_2) = \left(
                               \begin{array}{c}
                                 4 \\
                                -6
                               \end{array}
                             \right).
\end{align*}
We use $\alpha=(0.01, 0.02)^T$ and obtain $AV@R_\alpha^{reg}(X)=(-4, 20)^T+ \R^2_+$. Thus, the minimal risk compensating portfolio for a risk manager/regulator covers the worst case for the first and second asset which is $4$ units  and $-20$ units, respectively.  The reason is that $\alpha_i<p_n$ for all $i,n=1,2$ in this simple example which will reappear below.
\end{example}

\begin{example}
\label{ExAVaRReg_d=3}
Here $d=3$, $N=3$, $p = \of{1/3, 1/3, 1/3}$ and the payoff is given by
\begin{align*}
X(\omega_1) = \left(
                          \begin{array}{c}
                          4\\
                          3\\
                          1
                          \end{array}
                        \right), \quad X(\omega_2) = \left(
                               \begin{array}{c}
                                 6 \\
                                -5\\
                                -3
                               \end{array}
                             \right), \quad X(\omega_3) = \left(
                               \begin{array}{c}
                                 -2 \\
                                3\\ 
                                -4
                               \end{array}
                             \right).
\end{align*}
We use $\alpha_1 = \alpha_2 = \alpha_3 = 0.05$. If $M=\R^3$ (all initial portfolios are eligible) then $AV@R_\alpha^{reg}(X)=(2,5,4)^T + \R^3_+$,  and again the three worst cases are covered.

If $M$ is spanned by the two vectors $\of{5, 0, 1}^T$ and $\of{0, 10, 1}^T$, then $AV@R_\alpha^{reg}(X)$ has two vertices $(17.5, 5, 4)^T$ and $(2, 36, 4)^T$, and its recession cone is $M_+$ which is the convex cone generated by $\of{5, 0, 1}^T$ and $\of{0, 10, 1}^T$.

This example illustrates the feature indicated in remark \ref{RemInterpretationAVaRReg}: If $M$ is not of the form $\R^m \times \cb{0}^{d-m}$, then $AV@R_\alpha^{reg}(X)$ is in general not "point plus cone $M_+$".

One may interpret such an $M$ as follows: Deposits always involve the third asset which may serve as an insurance against the risks which come with the first two. For example, the first two assets may be currencies and the third an insurance against exchange risk or something like a (partial) gold exchange standard. In this sense, deposits in the third asset serve as a safeguard against exchange risk or simply inflation.
\end{example}

\begin{example}
Let $d = m = 2$, thus $M=\R^2$, $N=5$, $p=(0.25, 0.4, 0.3, 0.02, 0.03)^T$ and the following payoff
\begin{align*}
X(\omega_1) &= \left(
                          \begin{array}{c}
                          6\\
                          3
                          \end{array}
                        \right), \quad
X(\omega_2) = \left(
                               \begin{array}{c}
                                 -8 \\
                                -6
                               \end{array}
                             \right), \quad
X(\omega_3) = \left(
                          \begin{array}{c}
                          -4\\
                          2
                          \end{array}
                        \right), \\
X(\omega_4)& = \left(
                               \begin{array}{c}
                                 -90 \\
                                -6
                               \end{array}
                             \right), \quad
X(\omega_5) = \left(
                          \begin{array}{c}
                          -80\\
                          -60
                          \end{array}
                        \right).
\end{align*}
be given. With $\alpha=(0.05, 0.05)^T$ one obtains $AV@R_\alpha^{reg}(X)=(84.0, 38.4)^T + \R^2_+$. In contrast to the previous examples, the $\alpha$-levels do have an impact on the risk measure.
\end{example}

More complex examples with $d=5$ assets and $M = \R^2 \times \cb{0}^{3}$ and a random variable $X$ that is the payoff of an outperformance option can be found in examples~\ref{ex_d=5,m=2,r=0} - \ref{ex_d=5,m=2,r>0,tc_Bond} in the next section where  $AV@R^{reg}_\alpha$ as well as $AV@R^{mar}_\alpha$ are calculated.

\subsection{The discrete version of $AV@R_\alpha^{mar}$}

In this section, we will restrict ourselves to one-period models since market extensions including trading at times $t=0,...,T$ typically involve path dependent strategies in markets with transaction costs and should rather be formulated in a recursive way where attention should be paid to time consistency issues.

Let us assume from now on that trading is possible at time $t=0$ and $t=T$. We will use Benson's algorithm to calculate $AV@R_\alpha^{mar}\of{X}$ based on the ideas described in the previous subsection.

To do this, we need an analog of formula \eqref{AVaRRegDiscrete} for $AV@R_\alpha^{mar}\of{X}$. All conditions in \eqref{EqAVaRMarExt} are the same as in $AV@R_\alpha^{reg}$ except the condition $Z + X - z\One \in -C_T = K_0\One + L^0_d\of{K_T}$.
Let
\[
H = \left(
  \begin{array}{ccc}
    h^1_1 & \cdots & h^I_1 \\
    \vdots & \ddots & \vdots \\
    h^1_d & \cdots & h^I_d \\
  \end{array}
\right) \in  \R^{d \times I}
\]
be the matrix of generating vectors of $K_0$, see assumption (H3).

Under assumption (H5), we have that for each $\omega \in \Omega$ the cone $K_{T}\of{\omega}$ is spanned by $k^{1}\of{\omega}, \ldots, k^{J\of{\omega}}\of{\omega}$. Denote $J_n=J\of{\omega_n}$, $n=1, \ldots, N$, and let
\[
    \hat{J}=\sum_{n=1}^{N}J_n.
\]
Let
\[
A_{\of{n}} = \left(
              \begin{array}{ccc}
                k^1_{1n} & \cdots & k^{J_n}_{1n} \\
                \vdots &  & \vdots \\
                k^1_{dn} & \cdots & k^{J_n}_{dn} \\
              \end{array}
            \right) \in \R^{d \times J_n}, n=1,...,N
\]
be the matrices containing the generating vectors of $K_{T}\of{\omega}$ as columns and
\[
\hat{A} = \left(
            \begin{array}{cccc}
              A_{\of{1}} & 0 & \ldots & 0 \\
              0 & A_{\of{2}} &  & \vdots \\
              \vdots &  &   \ddots  & \\
               0&  & \ldots & A_{\of{N}} \\
            \end{array}
          \right) \in \R^{dN \times \hat{J}}
\]
a diagonal matrix where $0$ stands for blocks of zeros of appropriate dimensions. Consider the vector
\[
\hat s = \of{s_{11}, \ldots, s_{1J_1}, s_{21}, \ldots, s_{2J_2}, \ldots, s_{NJ_N}} \in \R^{\hat{J}}.
\]
Then the condition $Z + X - z\One \in L^0_d\of{K_T}+K_0\One$ can equivalently be written as
\[
\hat{z} +\hat{x} - \hat{I}_d z=\hat{A}\hat s+\hat{I}_dH\hat t
\]
for $\hat s \in \R^{\hat{J}}_+$ and $\hat t \in \R^I_+$.

Using this and the results of section~\ref{subsec AVAR_reg} for the other conditions we obtain
\begin{align}
\label{AVaRMarDiscrete}
AV@R_\alpha^{mar}\of{X} = \big\{ & \diag(\alpha)^{-1}\hat{P}\hat{z} - z \mid B_{\of{d-m}}(\diag(\alpha)^{-1}\hat{P}\hat{z} - z)=0,
\\
\nonumber
& \hat{z} +\hat{x} - \hat{I}_d z=\hat{A}\hat s + \hat{I}_dH \hat t, \; \hat{z}\in \R^{dN}_+, \; \hat s \in \R^{\hat{J}}_+, \; \hat t \in \R^I_+ , \; z \in \R^d\big\}.
\end{align}
Therefore, as in the regulator case, $AV@R_\alpha^{mar}\of{X}$ is the image of a polyhedral set under a linear function mapping $\R^{(N+1)d+I+\hat{J}}$ into $M \subseteq \R^d$. Its computation amounts to determine its minimal points with respect to the order generated by $K^M_0$.

Again, a few examples shall illustrate the theory for which we used an adapted version of the same algorithm. The input parameters are  the generating vectors $g^1, \ldots, g^L$ of the cone $K_0^M$ and the data from \eqref{AVaRMarDiscrete}.

\begin{example}
\label{ex_AVAR_mar_d=2}
Let all input parameters and the payoff $X$ be as in example \ref{ex_AVAR_reg_d=2}. Let the two assets be a USD cash account and one risky stock. The bid-ask prices of the risky stock at $t=0$ are given by $S_{0,b}=0.72,\; S_{0,a}=1$ and at $t=T$ by $S_{T,b}\of{\omega_1}=0.75,\; S_{T,a}\of{\omega_1}=1.11$ and  $S_{T,b}\of{\omega_2}=0.7,\; S_{T,a}\of{\omega_2}=0.9$. Then $AV@R^{mar}_\alpha\of{X}$ has two vertices given by $\of{-12, 20}^T$ and $\of{-39, 56}^T$ and a recession cone given by $K_0$.

With this result and  \eqref{AVaRMarDiscrete} in view one may ask the following question. What is the significance of the preimages which produce the two vertices? Two such preimages are the two points in $\R^{dN+\hat{J}+I+d} = \R^{4+4+2+2} = \R^{12}$ given by
\begin{align*}
\of{\hat{z}^1, \hat s^1, \hat t^1, z^1}^T & = \of{0,0,0,0,0,0,0,11.4286,0,0,12,-20}^T\\
\of{\hat{z}^2, \hat s^2, \hat t^2, z^2}^T & = \of{0,0,0,0,0,36,0,50,0,0,39,-56}^T
\end{align*}
which were also produced by Benson's algorithm. These two points are solutions of the problem represented by \eqref{AVaRMarDiscrete} in the sense that they generate two extremal minimal points of the set $AV@R_\alpha^{mar}\of{X}$, i.e., solutions of a linear vector optimization problem in a traditional sense. Moreover, it turns out that the components $\hat s^i, \hat t^i$, $i=1,2$, yield two trading strategies such that the corresponding payoffs $Y^1$, $Y^2$ satisfy
\[
AV@R^{mar}_\alpha\of{X} = \co\of{AV@R^{reg}_\alpha\of{X+Y^1}\cup AV@R^{reg}_\alpha\of{X+Y^2}}+K_0.
\]
A comparison with formula \eqref{EqAVaRInfRep} shows that the set $\cb{Y^1, Y^2}$ may be considered as a solution of the set-valued optimization problem established in remark \ref{RemMarExt2}. This corresponds precisely to the (mild) solution concepts given in \cite[Chap. 2]{Loehne11Book}.

The two trading strategies may be described as follows.  The first strategy consists of no trade at $t=0$ since $k_0^1 = H\hat t^1 = (0,0)^T \in K_0$, no trade at $t=T$ if $\omega_1$ occurs since $k_T^1(\omega_1) = A_1(\hat s_{11}^1, \ldots, \hat s_{1J_1}^1)^T = (0,0)^T \in K_T(\omega_1)$ and to sell $11.4286$ units of stock (at price $S_{T,b}\of{\omega_2}=0.7$ which yields $8$ USD) if $\omega_2$ occurs since $k_T^1(\omega_2) = A_2(\hat s_{11}^1, \ldots, \hat s_{1J_2}^1)^T = (-8,11.4286)^T \in K_T(\omega_2)$. If we denote by $Y^1\in C_T =-K_0\One - L^0_d\of{K_T}$ the outcome of this optimal strategy starting from zero capital at $t=0$ then $AV@R^{reg}_\alpha\of{X+Y^1} =(-12,20)^T + \R^2_+$ which corresponds to the first vertex of $AV@R^{mar}_\alpha\of{X}$.

The second strategy consists of no trade at $t=0$ since $k_0^2 = H\hat t^2 = (0,0)^T \in K_0$, to sell $36$ units of stock (at price $S_{T,b}\of{\omega_1} = 0.75$ which yields $27$ USD) if $\omega_1$ occurs  since $k_T^2(\omega_1) = A_1(\hat s_{11}^2, \ldots, \hat s_{1J_1}^2)^T = (-27,36)^T \in K_T(\omega_1)$ and to sell $50$ units of stock (at price $S_{T,b}\of{\omega_2} = 0.7$ which yields $35$ USD) if $\omega_2$ occurs since $k_T^2(\omega_2) = A_2(\hat s_{11}^2, \ldots, \hat s_{1J_2}^2)^T = (-35,50)^T\in K_T(\omega_2)$. If we denote by $Y^2$ the outcome of this optimal strategy starting from zero capital at $t=0$ then $AV@R^{reg}_\alpha\of{X+Y^2} = (-39,56)^T + \R^2_+$ which corresponds to the second vertex of $AV@R^{mar}_\alpha\of{X}$.

The two strategies lead to terminal payoffs $X + Y^1$ and $X + Y^2$, respectively, which can be considered as risk minimal with respect to the risk measure $AV@R^{reg}_\alpha$ among all possible terminal payoffs in $X + C_T$ according to remark \ref{RemMarExt2}. Since the representation \eqref{AVaRMarDiscrete} gives a linear relationship between points $\of{\hat{z}, \hat s, \hat t, z}$ and corresponding values $\diag(\alpha)^{-1}\hat{P}\hat{z} - z$ each convex combination of $Y^1$ and $Y^2$ also yields a risk minimal strategy with a minimal point of $AV@R_\alpha^{mar}\of{X}$ which is the corresponding convex combination of the two vertices above.

This shows that trading strategies which produce minimal elements of $AV@R_\alpha^{mar}\of{X}$ are also provided by Benson's algorithm.
\end{example}

\begin{remark}
\label{remark ex_set_vs_scalar}
Let us compare the set-valued approach to risk measurement to the scalar approach in which it is usually assumed that a multivariate random variable $X$ is first liquidated into a fixed num\'{e}raire asset and then a scalar risk measure is applied to the liquidated value of $X$ (see e.g. \cite{EkelandGalichonHenry10}, example 2.5, or the standard assumption of liquidation in the scalar approach to utility maximization in markets with transaction costs, see \cite{HodgesNeuberger89}).

In example~\ref{ex_AVAR_mar_d=2} the results would be as follows. The scalar $AV@R$ is calculated from the liquidated payoff, where we use the liquidation functions according to the bid-ask prices at time $T$
\begin{align*}
    l_1\of{X}\of{\omega}&=X_1\of{\omega}+X_2\of{\omega}S_{T,b}\of{\omega}I_{\{X_2\of{\omega}\geq 0\}}+X_2\of{\omega}S_{T,a}\of{\omega}I_{\{X_2\of{\omega}< 0\}},\\
    l_2\of{X}\of{\omega}&=X_2\of{\omega}+\frac{X_1\of{\omega}}{S_{T,a}\of{\omega}}I_{\{X_1\of{\omega}\geq 0\}}+\frac{X_1\of{\omega}}{S_{T,b}\of{\omega}}I_{\{X_1\of{\omega}< 0\}}.
\end{align*}
In example~\ref{ex_AVAR_mar_d=2} the liquidated payoff is $l_1\of{X}=\of{-10.2,-1.4}$ if $X$ is liquidated into the first asset, and $l_2\of{X}=\of{-9.19,-1.6}$ if $X$ is liquidated into the second asset. One obtains $AV@R^{sca}_{\alpha_1}\of{l_1\of{X}} = 10.2$ USD and $AV@R^{sca}_{\alpha_2}\of{l_2\of{X}} = 9.19$ units of stock.

The set $AV@R^{mar}_\alpha\of{X}$ from example~\ref{ex_AVAR_mar_d=2} intersects the $1$- and $2$-axis at $\of{8,0}$ and $\of{0,8}$, respectively. Clearly,  $\of{AV@R^{sca}_{\alpha_1}\of{l_1\of{X}}, AV@R^{sca}_{\alpha_2}\of{0}} = \of{10.2, 0}$ and 
$\of{AV@R^{sca}_{\alpha_1}\of{0}, AV@R^{sca}_{\alpha_2}\of{l_2\of{X}}} = \of{0, 9.19}$ are points in the interior of $AV@R^{mar}_\alpha\of{X}$.  This perfectly corresponds to findings in \cite{ArtDelKochMed09} where it is shown that a multi-eligible asset approach may decrease the costs of superhedging.
\end{remark}

\begin{example}
\label{ex_d=5,m=2,r=0}
We consider an example with $d=5$ assets. The first asset $S^1$ is a USD cash account (zero interest rate), the second is another currency (e.g. EUR), denoted in USD, and the other assets are risky stocks denoted in USD. As the space of eligible assets we choose the space spanned by the first and the second asset (the currencies), i.e. $M = \R^2 \times \cb{0}^{3}$.
Let the initial prices of the $5$ assets in USD be given by $S_0=(1, 1.3, 50, 6, 25)^T$, let the covariance matrix of the $4$ risky assets be
 \[
 \left(
        \begin{array}{cccc}
         0.010  &  0.004 &    0.002 &   0.018\\
    0.004 &  0.040&    0.012&    0.006\\
    0.002&    0.012&    0.0225&    0.012\\
    0.018&    0.006&    0.012&    0.040
        \end{array}
      \right).
 \]
and $\mu=(0.03, 0.1, 0.06, 0.12)^T$ the drift vector of the $4$ risky assets. Let the length of the one-period model under consideration be one year.

We will follow the method in \cite{KornMueller09} to set up a (one-period) tree for the risky assets that reflects the drift and correlation structure using the decoupling method with the Cholesky decomposition, see \cite{KornMueller09} and also section~5.2 in \cite{LR11} for more details in a setting similar to ours in markets with transaction costs. We adapt the method to obtain a tree under the real world probability measure. The one-period tree will have $2^{d-1}=16$ branches, i.e. $N=16$ in this example and the probabilities of each path are given by $2^{-4}$.
 Now, let us assume that the proportional transaction costs for the risky assets are given by $\lambda=(\lambda^2,...,\lambda^{d})^T=(0.07, 0.05, 0.01, 0.01)^T$ and that the bid and ask prices at $t\in\{0,T\}$ are given by
\begin{align*}
    (S_{t}^b)^i=S_{t}^i(1-\lambda^i),\quad\quad (S_{t}^a)^i=S_{t}^i(1+\lambda^i),\quad \quad i=2,...,d.
\end{align*}
Furthermore, let us assume an exchange between any two risky assets can not be made directly, only via cash in USD by selling one asset and buying the other. Since all
risky assets are denoted in USD, the solvency cone $K_t$ for $t\in\{0,T\}$ is generated by the columns of the following matrix (see e.g. \cite{LoehneRudloff_solv_working})
\begin{align}
\label{solvcone}
\left(
  \begin{array}{cccccccc}
    (S_{t}^a)^2 & -(S_{t}^b)^2 &(S_{t}^a)^3 & -(S_{t}^b)^3&(S_{t}^a)^4 & -(S_{t}^b)^4 &(S_{t}^a)^5 & -(S_{t}^b)^5\\
    -1 & 1 & 0 & 0& 0 & 0& 0 & 0\\
     0 & 0&-1 & 1 & 0 & 0& 0 & 0\\
     0 & 0& 0 & 0&-1 & 1 & 0 & 0\\
     0 & 0& 0 & 0& 0 & 0 & -1 & 1
  \end{array}
\right).
\end{align}
As the payoff $X$ we consider an outperformance option with strike $K=(1+\lambda_1)S_0^2= 1.378$ and physical delivery. This option gives the right to buy the asset that performed best out of a basket of assets at a given strike price. A vector $c$ is defined as $(S_0^a)^2=c_i(S_0^a)^i$ for $i\in\{2,...,5\}$. Let the payoff $X$ be $-K$ in the USD account, $c_i$ units of asset $i$ for the smallest $i$ satisfying $c_i(S_T^a)^i=\max_{j\in\{2,...,5\}}(c_j(S_T^a)^j)\geq K$ and zero in the other assets. If $\max_{j\in\{2,...,5\}}(c_j(S_T^a)^j)< K$ the payoff is the zero vector.
The maturity $T$ is chosen as one year.

Let us calculate $AV@R^{reg}_\alpha\of{X}$ and $AV@R^{mar}_\alpha\of{X}$ with $\alpha=(0.1, 0.08, 0.09, 0.1, 0.05)^T$.
Formula \eqref{AVaRRegDiscrete} leads to a problem with $85$ variables, $166$ constraints and $2$ objectives. Formula \eqref{AVaRMarDiscrete} for $AV@R^{mar}\of{X}$ leads to a problem with $221$ variables, $302$ constraints and $2$ objectives.
$AV@R^{reg}_\alpha\of{X}$ has one vertex at  $(1.3910, 0)^T$, which is the smallest cash deposit in the first two assets necessary to compensate the risk of $X$ without involving trading. The recession cone of $AV@R^{reg}_\alpha\of{X}$ is $\R^2_+$.
The set $AV@R^{mar}\of{X}$ has five vertices that are given by the columns of the following matrix
 \begin{align*}
\left(
  \begin{array}{ccccc}
    0.8858 & 0.4200 & 0.3404 & 0.2794& -0.0699 \\
     -0.7160 & -0.3771&-0.3166 & -0.2698 & 0 
  \end{array}
  \right)
\end{align*}
and the recession cone $K_0^M$.
\end{example}

\begin{example}
\label{ex_d=5,m=2,r>0}
Let us consider the same model as in example~\ref{ex_d=5,m=2,r=0}, now with an annual interest rates of $5\%$ for the riskless asset denoted USD. Let $(S_0)^1=(1+r)^{-1}=0.9524$ and $(S_T)^1=1$. All the other input parameters are as before. The solvency cones change in the sense that in the matrix \eqref{solvcone} all the values $\pm 1$ are replaced by $\pm (S_0)^1$, see \cite{LoehneRudloff_solv_working}. 

The set $AV@R^{reg}_\alpha\of{X}$ has one vertex at  $(1.391, 0)^T$ and recession cone $\R^2_+$, i.e. it is the same as before.  To calculate the smallest deposit in cash (USD and EUR), one just needs to multiply the number of USD bonds with the initial bond price $(S_0)^1=0.9524$. 

The set $AV@R^{mar}_\alpha\of{X}$ has, in contrast to the previous case, seven vertices given by the columns of the following matrix
 \begin{align*}
\left(
  \begin{array}{ccccccc}
    4.7411 & 4.1666 & 3.9181 & 0.8716&  0.3954 	&0.2508	& -0.1087 \\
     -3.3883 & -2.9941&-2.8235 &-0.7160 &  -0.3771	&-0.2698	& 0
  \end{array}
  \right)
\end{align*}
and the recession cone $K_0^M$.
\end{example}

\begin{example}
\label{ex_d=5,m=2,r>0,tc_Bond}
Let us consider the same model as in example~\ref{ex_d=5,m=2,r>0}, now with transaction costs $\lambda_0=0.03$ for the riskless asset. That means $(S_0^b)^1=(1-\lambda_0)(1+r)^{-1}$, $(S_0^a)^1=(1+\lambda_0)(1+r)^{-1}$ and $(S_T^b)^1=(1-\lambda_0)$, $(S_T^a)^1=(1+\lambda_0)$. All the other input parameters are as before. The solvency cones now have $20$ generating vectors instead of $8$, for details see e.g. \cite{LoehneRudloff_solv_working}.

Again, $AV@R^{reg}_\alpha\of{X} = (1.391, 0)^T + \R^2_+$. The set $AV@R^{mar}_\alpha\of{X}$ has eleven vertices (in USD bonds and EUR) given by the columns of the following matrix
  \begin{align*}
\left(
  \begin{array}{ccccccccccc}
    -0.0640& 	6.9490 &	5.0535 & 	4.2335&	1.8394 & 	1.3989 &	1.1107 & 1.0965& 0.4243 &   0.2970  &  0.2024\\
     0&		-4.7770 & -3.5156 & -2.9696 & -1.3630& -1.0652 & -0.8621& -0.8520&  -0.3700 &  -0.2746&   -0.2033
  \end{array}
\right)
\end{align*}
and the recession cone $K_0^M$. Formula \eqref{AVaRMarDiscrete} for $AV@R^{mar}_\alpha\of{X}$ leads to a problem with $425$ variables, $506$ constraints and $2$ objectives.
\end{example}

\begin{example}
Let us calculate the market extension $AV@R^{mar}_\alpha\of{X}$ for the payoff  $X$ and all parameters given by example~\ref{ExAVaRReg_d=3}, where $M$ is spanned by the two vectors $\of{5, 0, 1}^T$ and $\of{0, 10, 1}^T$. Let the $d=3$ assets be the first three assets of example~\ref{ex_d=5,m=2,r=0}. The one-period tree has $2^{d-1}=4$ branches, i.e. $N=4$. Then, the set $AV@R^{mar}_\alpha\of{X}$ has five vertices  given by the columns of the following matrix
  \begin{align*}
\left(
  \begin{array}{ccccc}
   -1153.3   &-656.2 &  -544.0    &13.5&    118.6\\
    2056.4 &   1172.2 &   973.4 &   5.0&   -177.1\\
   -25.0&   -14.0 &  -11.5&    3.2&    6.0
  \end{array}
\right)
\end{align*}
and the recession cone $K_0^M$.
\end{example}

{\bf Acknowlegdment.} The authors emphasize that this project benefited from discussions with Frank Heyde (about the primal definition of AV@R) 
and Andreas L\"ohne (about the computational part). We are grateful to the referees of a previous version for their constructive remarks.

\bibliographystyle{spmpsci}      
\bibliography{databasekl}   

\end{document}